\documentclass[11pt,twoside]{article}
\usepackage{asp2010}

\resetcounters

\begin{document}

\title{Identifying the chemistry of the dust around AGB stars in nearby galaxies}
\author{S. Srinivasan$^1$, F. Kemper$^1$, and R. Zhao-Geisler$^{2,1}$
\affil{$^1$Institute of Astronomy and Astrophysics, Academia Sinica, Taiwan}
\affil{$^2$National Taiwan Normal University, Taipei, Taiwan}
}

\begin{abstract}
Asymptotic giant branch (AGB) stars are significant contributors to the chemical enrichment of the interstellar medium (ISM) of galaxies. It is therefore essential to constrain the AGB contribution to the dust budget in galaxies. Recent estimates of the total AGB dust injection rate to the Large and Small Magellanic Clouds \citep[LMC and SMC;][Srinivasan et al. in prep]{Riebeletal2012,Boyeretal2012} have used data from the Spitzer Space Telescope SAGE \citep[Surveying the Agents of Galaxy Evolution;][]{Meixneretal2006} and SAGE-SMC \citep[][]{Gordonetal2011} surveys. When sorted by dust chemistry, the data allow for a comparison of O--rich and carbonaceous dust production rates. In the LMC, for instance, the rate of dust production from carbon stars is two and a half times that from oxygen-rich AGBs. A reliable determination of the fractional contributions of the two types of dust would serve as input to models of chemical evolution. However, the Spitzer IRAC photometric bands do not sufficiently probe the characteristic mid-infrared spectral features that can distinguish O--rich AGBs from carbon stars -- namely, the 9.7 $\mu$m silicate feature and 11.3 $\mu$m silicon carbide feature.\\
With the continuous spectral coverage in the ~4-30 $\mu$m range, SPICA has the potential to distinguish these two types of chemistries. In this contribution, synthetic photometry from the model grid of AGB stars, GRAMS \citep{Sargentetal2011, Srinivasanetal2011}, will be used to discuss the science possibilities that SPICA might offer this study.
\end{abstract}

\section{The AGB dust budget in nearby galaxies}
Low- and intermediate-mass (0.8 -- 8 M$_\odot$) stars go through the asymptotic giant branch (AGB) stage towards the end of their lives. During this phase of the star's lifetime, the products of shell hydrogen and helium burning are mixed into the outer layers by huge convective zones. Surface pulsations levitate the material to cooler regions where they form gas molecules and, further on, solid particles (dust). Radiation pressure on the dust grains then drives an efficient outflow \citep[{\it e.g.},][]{Woitke2006,BladhHofner2012}. The mass-loss rate increases as the star evolves along the AGB, and can exceed $10^{-4}$ M$_\odot$ yr$^{-1}$. Stars can thus lose a significant fraction of their mass, in the form of gas and dust, during the AGB phase. The ejecta are mixed into the interstellar medium (ISM) where the dust can be incorporated into the next generation of stars. By affecting the subsequent star formation while simultaneously increasing the metal content in the ISM, the amount of AGB mass loss provides an important constraint to galactic chemical evolution and population synthesis models.\\

Perhaps the most important constituent of the dredged-up matter is $^{12}$C, which quickly combines with any existing oxygen atoms to form CO. The subsequent chemistry of the circumstellar shell is dependent on the slight overabundance of either O or C. With progressive dredge-up episodes, the abundance of carbon relative to oxygen (the C/O ratio) exceeds unity leading to the formation of carbon stars for initial masses of up to $\sim$ 4 M$_\odot$, depending on the metallicity. Given their mass range, carbon stars are numerous at low metallicities; however, individual massive O--rich AGBs have large dust production rates (DPRs) by virtue of their mass. In order to estimate the relative contributions of carbonaceous material and silicate/metal oxides, it is necessary to sort the AGB population based on circumstellar chemistry. While it is easy to identify optically thin O--rich and C--rich AGBs from their near-infrared (NIR) colours, the identification of dusty sources (the so-called ``extreme" AGBs) requires mid-infrared data \citep[see, {\it e.g.}, Figure 5 in][]{Boyeretal2011}. Even with this information, it is not straightforward to determine the circumstellar chemistry -- the extreme AGB population consists of a mixture of O--rich AGBs and carbon stars. Based on their masses, it is expected that most of these stars are carbon-rich; however, even a handful of massive O--rich extremes in this population can contribute significantly to the dust budget. The chemistry of the dust can be confirmed with mid-infrared (MIR) spectra owing to the typical prominent spectral signatures of AGB dust -- the silicate feature at 9.7 $\mu$m in O--rich stars, and the silicon carbide feature at 11.3 $\mu$m for carbon stars. In practice, such spectra are available only for a small fraction of the AGB population; estimates of the integrated dust budget rely largely on photometric data.\\

Studies of Galactic AGB stars are affected by moderate to high line-of-sight extinction, resulting in uncertain distance and luminosity estimates. Among nearby galaxies, the Magellanic Clouds are ideal for AGB studies because of their proximity, viewing angle and low extinction along their lines of sight. The Clouds were imaged as part of the {\it Spitzer} Space Telescope SAGE \citep{Meixneretal2006} and SAGE-SMC \citep{Gordonetal2011} surveys. The SAGE-Spec program \citep[][]{Kemperetal2010} obtained follow-up MIR spectra for a subset of SAGE point sources. The evolved-star candidates extracted from these data were used to calculate the dust budget based on empirical relations between the DPR and broadband MIR properties such as the colour \citep{Matsuuraetal2009, Matsuuraetal2013} and the excess flux from dust \citep{Srinivasanetal2009, Boyeretal2012}. While such empirical relationships enable quick estimates for the cumulative DPR, they assume that the trends determined from a handful of well-studied sources apply over the entire range of spectral energy distributions (SEDs) of the evolved star population.\\

While radiative transfer modelling of the individual SEDs, taking into account all the available information \citep[{\it i.e.}, spectra along with preferably multi-epoch photometry to constrain the variability of the sources; see, {\it e.g.},][]{Sargentetal2010,Srinivasanetal2010}, provides more accurate estimates of the DPR, such individual modelling is unrealistic for large datasets. As a compromise, SEDs could be fit to a pre-computed grid of radiative transfer models that spans the range of observed parameters for dusty evolved stars. This was the motivation for the {\bf G}rid of {\bf R}ed supergiant and {\bf A}GB {\bf M}odel{\bf S} \citep[GRAMS;][]{Sargentetal2011,Srinivasanetal2011}. The GRAMS grid contains over 66\,000 models with silicate dust and about 12\,000 models with a mixture of amorphous carbon and 10\% SiC by mass, and the models reproduce the range of observed MIR properties of LMC AGB and red supergiant (RSG) stars. \citep{Riebeletal2012} used the GRAMS grid to fit the SEDs of LMC evolved star candidates. They found that (a) the total dust return from AGB and RSG stars to the ISM is about $2.1\times 10^{-5}$ M$_\odot$ yr$^{-1}$, (b) the return from carbon stars is 2.5 times that from O--rich AGBs and RSGs, (c) carbon stars comprise more than 95\% of the extreme AGB population, and (d) although the extreme AGBs amount for only $\leq$5\% of the evolved star population, they contribute $\geq$75\% of the dust return. A similar study for the SMC is in progress (Srinivasan et al. in prep).

\section{AGB science with SPICA}
As previously mentioned, it is important to classify the evolved star candidates by chemical type in order to determine the relative contributions of carbonaceous and oxygen-rich dust. The main hindrance to MIR colour-based classifications of {\it Spitzer} datasets is the absence of filters that can efficiently probe the 9.7 $\mu$m silicate in O--rich AGBs and the 11.3 $\mu$m SiC features in carbon stars. While these features may contribute slightly to the IRAC 8 $\mu$m band (similarly, the silicate feature at $\sim$18 $\mu$m may affect the flux in the MIPS 24~$\mu$m band), it is not easy to constrain the dust parameters such as optical depth and DPR with just these bands. If available, {\it Spitzer} data can be complemented with either AKARI or WISE observations, both of which offer bands in the $\sim$11--12 $\mu$m range. The continuous coverage over the 0.7--38 $\mu$m range offered by the Focal Plane Camera (FPC) and Mid-Infrared Camera and Spectrograph (MCS) instruments on board SPICA is ideal for AGB-star studies.

\subsection{Customising the SPICA MCS for AGB chemical classification}
The specification for the SPICA MCS-S filter set consists of ten filters covering wavelengths from about 3.3 $\mu$m to 24~$\mu$m (see Sakon, these proceedings, for more details on these filters). This current configuration is well-suited for identifying most AGB features in the MIR. In addition, the five filters on the FPC can be used to constrain the photospheric contribution to the SED.\\

\begin{figure}[!ht]
\begin{center}
   \resizebox{0.8\hsize}{!}{
     \includegraphics*{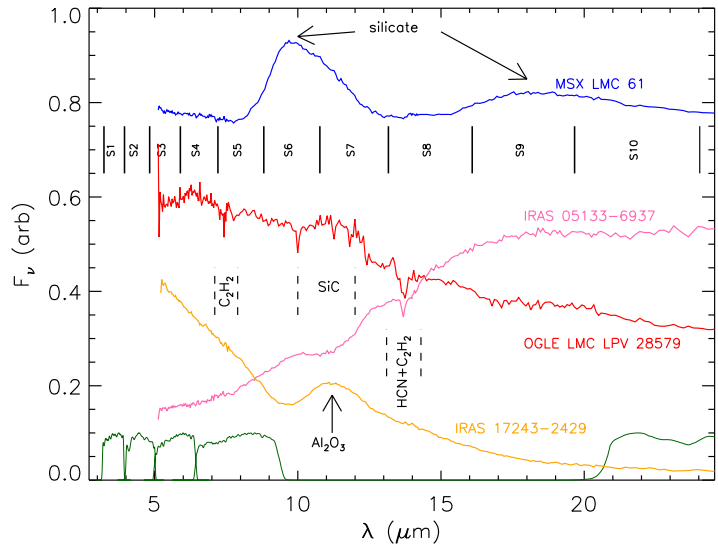}
   }
\end{center}
\caption{The SAGE-Spec spectra of an O--rich AGB (blue) with silicate emission and two carbon stars with SiC in emission (red) and in absorption (pink), and the spectrum of an S-type star from \citet{Smoldersetal2012} with strong alumina emission (orange). The vertical dashed lines show the approximate location and extent of the SiC feature and molecular absorption due to acetylene and HCN in carbon stars. The filter response curves for the IRAC and MIPS 24~$\mu$m bands (green) are compared to the SPICA MCS-S filter set (vertical solid lines). The continuous coverage of this wavelength range by the SPICA MCS-S filters (solid vertical lines), unlike the IRAC and MIPS~24~$\mu$m filters (filter response curves in green), recovers the silicate as well as the SiC and alumina features. In addition, the SPICA bands are also able to detect strong C$_2$H$_2$ absorption at 13.7 $\mu$m.
}\label{fig:comparespica}
\end{figure}
\begin{figure}[!ht]
\begin{center}
   \plottwo{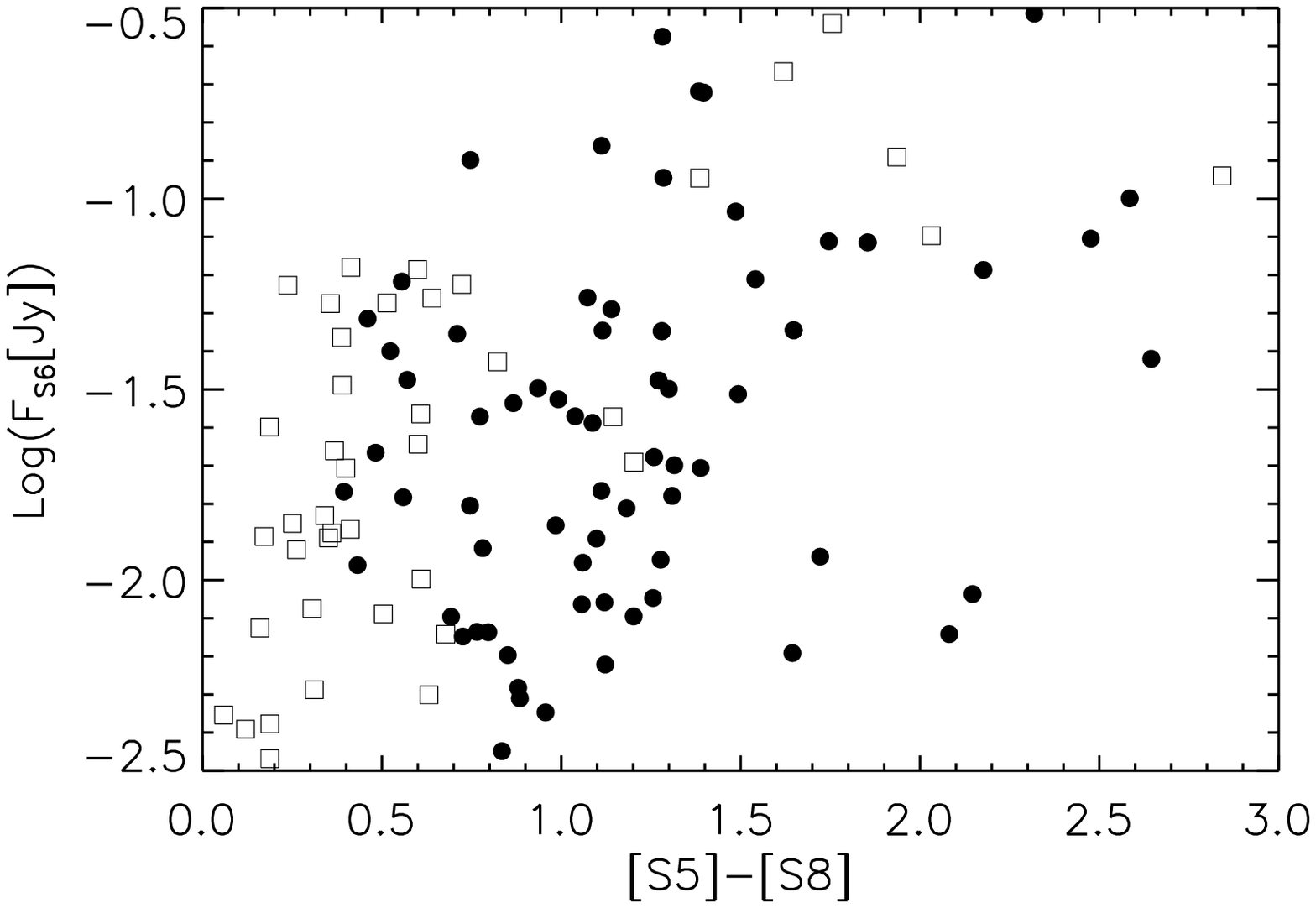}{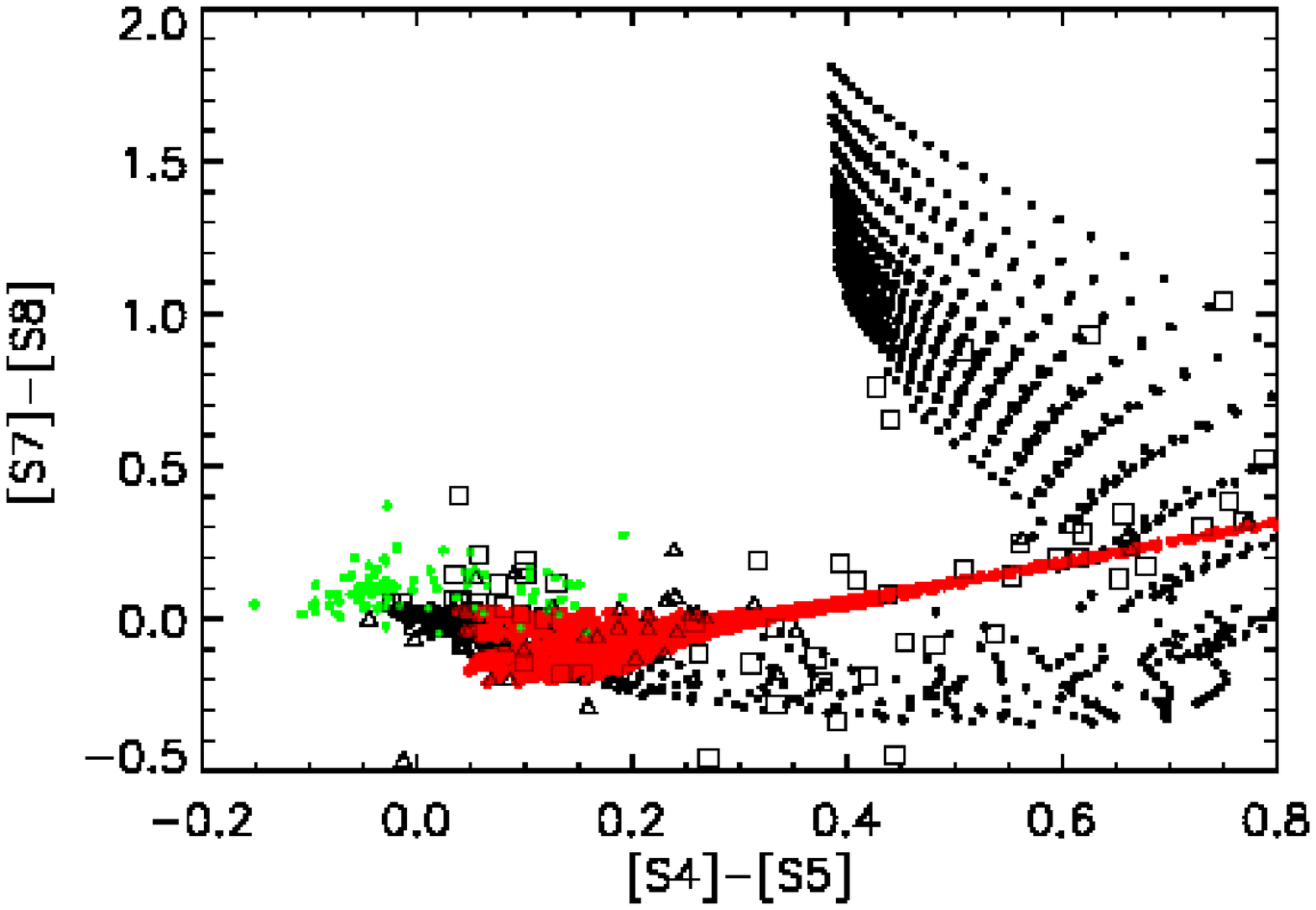}
\end{center}
\caption{{\it Left:}\, SAGE-Spec O-rich AGBs with 9.7 $\mu$m silicate emission (circles) are well separated from carbon stars (squares) in a plot showing the flux in the S6 filter versus the [S5]--[S8] colour. The flux in the S6 filter is proportional to the silicate feature strength, and the colour is a measure of the pseudo-continuum around the feature. {\it Right:}\, The \citet{Smoldersetal2012} S-type stars (green circles) occupy a small region on this SPICA MCS colour-colour diagram. SAGE-Spec carbon stars (triangles) and O--rich AGB stars with silicate emission (squares) as well as the GRAMS O--rich (black circles) and C--rich (red) models are also shown for comparison.}\label{fig:f2cccd}
\end{figure}

Figure~\ref{fig:comparespica} compares the {\it Spitzer} IRS spectra of different types of AGB stars. The silicate features at 9.7 $\mu$m and 18 $\mu$m, while outside the range of the IRAC and MIPS 24~$\mu$m filters, are covered by the MCS filter set. The 11.3 $\mu$m SiC feature straddles two filters. Moreover, it is difficult to estimate the continuum in LMC carbon stars around the SiC feature due to molecular absorption features ({\it e.g.}, the strong C$_2$H$_2$ feature at 13.7 $\mu$m). For this reason, it is recommended that the MCS include a narrow-band filter centred around 11.2 $\mu$m. Despite this difficulty, the SPICA MCS filter set can be used to separate O--rich and C--rich features in AGB stars. This is illustrated in Figure~\ref{fig:f2cccd} (left panel), where the flux in the S6 filter (centred at $\sim$9.7 $\mu$m, see Figure \ref{fig:comparespica}) is plotted against the [S5]--[S8] colour, which is a measure of the continuum around the silicate feature for O--rich AGBs. The SAGE-Spec sources with the 9.7 $\mu$m silicate feature in emission separate out from the carbon stars on this diagram.\\

%

In addition to O-- and C--rich chemistries, it may also be possible to identify regions in colour-colour space that occupied by S-type AGB stars -- sources with C/O ratios close to unity. To demonstrate this, photometry in the SPICA bands synthesised from IRS spectra of galactic S-type stars from the \citet{Smoldersetal2012} sample is compared with the GRAMS models in Figure~\ref{fig:f2cccd} (right panel). S-type stars and galactic O--rich AGBs also feature broad Al$_2$O$_3$ emission at $\sim$11.2 $\mu$m; as with the SiC feature, the alumina feature is split across two adjacent MCS filters in the current configuration (Figure \ref{fig:comparespica}). A narrow-band filter would therefore be useful for both features.

\subsection{Circumstellar molecules}
Dusty AGB stars in low-metallicity populations such as the LMC show prominent molecular absorption features in the MIR. These features are attributed to a molecular layer in the circumstellar shell and, in principle, they can be used to find the column density and gas:dust ratio in these sources \citep[see, {\it e.g.},][]{Matsuuraetal2006,Sargentetal2010,Srinivasanetal2010}. However, the $\sim$4.4--6 $\mu$m CO band is at the edge of the {\it Spitzer} IRS spectrum, which hampers accurate parameter estimates. The band centred at $\sim$4.4 $\mu$m in the MCS-S filter set may be able to provide a rough estimate of the CO feature strength in AGB stars. In addition, the features due to other molecules such as HCN and C$_2$H$_2$ can be captured by filters centred at 8.0 $\mu$m and 14.6 $\mu$m  (Figure~\ref{fig:comparespica}).\\

The SPICA MCS filters will help in the chemical classification of AGB stars, even when such classification is difficult based on near-infrared colours. The filters will also provide tighter constraints on the optical depth and dust-production rates by targeting prominent AGB dust features.  The current specification for the filters works well for this purpose, but a narrow-band filter around 11.2 $\mu$m would be a great addition to this set for the broad silicon carbide and alumina features centred in this wavelength range. The SPICA MCS filter set, along with the FPC, offers a great opportunity for galaxy-wide studies of AGB populations in nearby galaxies (see, {\it e.g.}, Zhao-Geisler, these proceedings).

\bibliography{lSPICA}

\end{document}